# Perspective on descriptors of mechanical behavior of cubic transition-metal carbides and nitrides

Hanna Kindlund,[1*] Theodora Ciobanu,[2] Suneel Kodambaka,[1] and Cristian V. Ciobanu[3]

[1]Department of Materials Science and Engineering,

Virginia Polytechnic Institute and State University, Blacksburg, VA 24061, USA

[2]University of Notre Dame, College of Arts and Letters, Notre Dame, IN 46556, USA

[3]Department of Mechanical Engineering and Materials Science Program,

Colorado School of Mines, Golden CO 80401, USA

**Abstract**

Cubic rocksalt structured transition-metal carbides, nitrides (TMC/Ns), and related alloys, are attractive for a wide variety of applications, notably as hard, wear-resistant materials. To-date, valence electron concentration (VEC) is used as a good indicator of stability and mechanical properties of these refractory compounds. In this perspective, we argue for the need of electronic descriptors beyond VEC to explain and predict the mechanical behavior of the cubic TMC/Ns. As such, we point out that descriptors that highlight differences between constituents, along with semi-empirical models of mechanical properties, have been underused. Additionally, it appears promising to partition VEC into contribution to ionic, covalent, and metallic bonds and we suggest that such partition could provide more insights into predicting mechanical properties in the future.

[*] Corresponding authors: kindlund@vt.edu

Transition-metal carbides and nitrides (TMC/Ns), especially those made of group 4, 5, and 6 elements, possess remarkable properties: they are some of the hardest (tens of GPa), high moduli (>300 GPa), and highest melting point solids (see Table I) with high-temperature mechanical strengths, high electrical and thermal conductivities, and excellent wear, ablation, and corrosion resistance.[1-4] These hard refractory ceramics, owing to a mixture of ionic, covalent, and metallic bonding, form a technologically important class of materials widely used as protective coatings on cutting tools[5] and as structural components in aerospace vehicles, hypersonic jets, and other systems operating in extreme environments.[6-8] These materials are also of interest as catalysts,[9, 10] energy storage materials,[11] and as metallic interconnects and diffusion barriers in electronics.[12] Continued research in this class of materials stems from the necessity to discover and develop new materials with improved mechanical properties for a variety of structural applications.

| Material | Melting point [$^o$C] | $E$ [GPa] | $H$ [GPa] |
|---|---|---|---|
| TiC (TiN) | 3067 (2949) | 431 (463) | 24 (25) |
| ZrC (ZrN) | 3420 (2982) | 444 (403) | 27 (19) |
| HfC (HfN) | 3928 (3387) | 395 (392) | 24 (20) |
| VC (VN) | 2648 (2177) | 520 (406) | 27 (15) |
| NbC | 3600 (--) | 504 (335) | 26 (10) |
| TaC (TaN) | 3983 (--) | 533 (319) | 26 (8) |
| VMoN | -- | 376 | 21 |
| VWN | -- | 350 | 23 |

**Table I.** Melting point, elastic modulus E, and hardness H of a few selected TMC/Ns, from Refs.[3, 4, 13, 14]

Alloying is probably the most commonly used approach to tailor the mechanical properties of materials. Classical examples of alloying include gold jewelry, bronze, brass, and stainless steel. The role of alloying elements on mechanical behavior of metallic materials,[15] as well as simple descriptors of structural stability, such as the valence electron concentration (VEC), have been widely reported in the literature.[16, 17] The VEC, a simple but fundamental parameter, describes well the effects of alloying on the structure and mechanical properties in metals. The insights gained from these studies on metallic alloys are often transferred to ceramic materials,[18, 19] even though the chemical bonding and crystal structures, that are fundamental parameters behind the mechanical behavior of materials, are significantly different in these compounds compared to metals. The effect of alloying on mechanical properties of TMN/Cs is still not well understood since *systematic studies* are scarce in comparison to those on metals. Therefore, our ability to predict and design refractory TMC/Ns alloys with desired mechanical properties has been rather limited. Only recently, Brenner, Curtarolo, and co-workers[20-27] have developed strategies for predicting the stability and properties of multicomponent, high-entropy TMC alloys.

Over the past decades, a major research goal has been to enhance the hardness of protective coatings for various applications. Among the TMC/Ns, probably group 4 and group 4 based alloys (e.g., TiAlN) have been the most extensively studied.[28-32] In search of ultra-hard coatings, early studies[5, 6, 32-35] focused on the incorporation of smaller atoms (B, Al, and Si) into the TMC/N lattices to promote stronger bonding and, hence, further enhance the mechanical strength of these ceramics. However, an increase in hardness is also followed by an increase in brittleness. Therefore, while high hardness is indispensable, focusing in increasing it alone at the expense of decreasing the ductility is not sufficient for most applications. For example, in the cutting tool and aerospace industries, where the materials are exposed to high thermo-mechanical stresses, increasing the extent of plastic deformation upon yielding (in addition to strength) is essential to avoid brittle failure. Therefore, in many applications, high modulus, high strength, *and* large plastic strain materials are required to increase the lifetime of any structural component. Recent efforts have aimed at enhancing the ductility of these refractory compounds[36-39] leading to the fabrication of tough ceramics, for example, by the incorporation of ductile phases, nanoscale grains, and multilayered structures.[37, 40, 41] *However, most of the methods for improving the ductility of hard ceramics are based on trial-and-error approaches with very few reports focusing on the electronic origins of hardness vs ductility*.[18, 19, 29, 42-53] This aspect of realizing both high strength and large plastic strain ceramics is a long-standing challenge. Therefore, despite their exceptional strength, the use of TMC/Ns coatings, for instance, in low-temperature structural applications has been rather limited. If we can increase their ductility while retaining their high strength, their application potential could be improved. Thus, gaining insight into the effect of alloying at the electronic level is crucial to design and predict hard-yet-tough ceramics.

For the B1-structured TMC/Ns, existing reports[21,39][13, 14, 54] have already provided some guidelines on how mechanical stability and properties such as moduli, hardness, and shear strength are expected to vary with the electron density in the $d$-$t_{2g}$ orbitals. In these carbides and nitrides, the strong $p$(N,C)-$d$-$e_g$(metal) first-neighbor bonds are responsible for the material's strength, while the relatively weaker metallic metal-metal $d$-$t_{2g}$ second-neighbor interactions control the material's ductility. Since the $d$-$t_{2g}$ orbital occupancy is related to the VEC, most existing studies have used VEC to compare and contrast the mechanical properties of different TMC/Ns.[4, 19, 42, 44] Holleck[6] and Jhi *et al.*[19, 55, 56] reported that maximum hardness in cubic TMC/Ns is achieved at a VEC of ~8.4 electrons, due to complete filling of the shear-resistive $p$-$d_{eg}$ orbitals. At a higher VEC, the shear sensitive $d$-$t_{2g}$ orbitals begin to be filled reducing the shear-resistance of the material and hence reducing its hardness. Sangiovanni *et al.*[44, 57] and others[42, 58] also used a similar approach to predict toughness enhancement with increasing the VEC. For clarity, we state that the ductility region is often specified by a Pugh's ratio $G/B$ below 0.5 *and* a Poisson ratio $\nu$ above 0.28.[4] Following these predictions, Kindlund and co-workers [14, 54] carried out a DFT-inspired experimental study and investigated the mechanical behavior of pseudobinary B1-structured group 5+6 transition-metal alloy nitrides, VMoN and VWN.[13, 59-61] They demonstrated that the introduction of group 6 elements (Mo and W) in group 5 nitride (VN) increases the ductility, while retaining the strength, i.e. these alloys exhibited enhanced toughness. VMoN (with VEC ~ 10.5) is tougher than the parent binary compound VN and the reference group 4 nitride, TiN VEC = 9). These pioneering experiments laid the foundation for rational design of refractory TMC/Ns with the most sought-after mechanical property in this class of materials, toughness –i.e., the combination of high strength and high ductility.

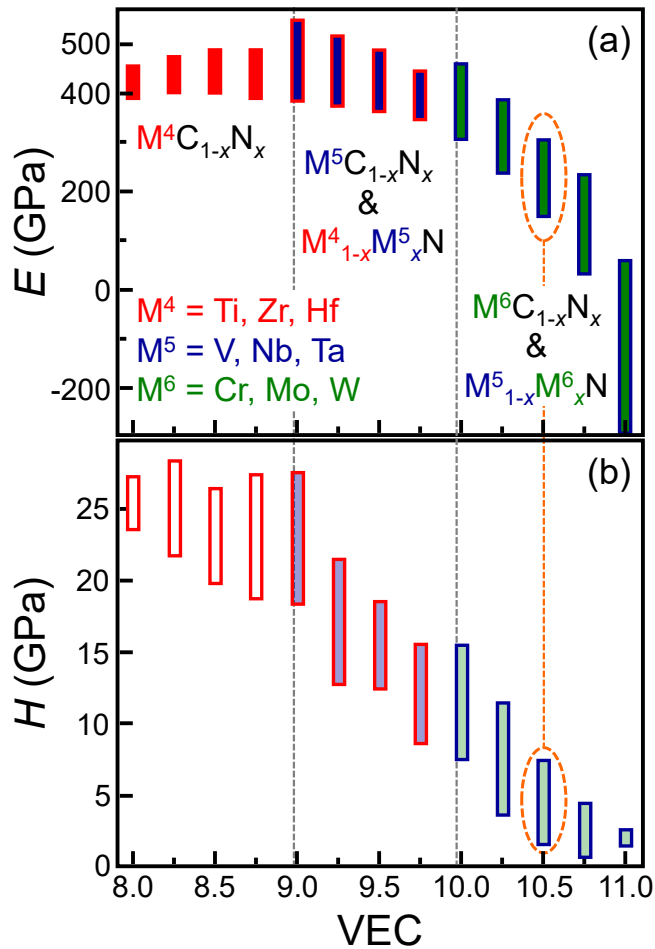

**Figure 1.** Trends in DFT-calculated elastic modulus $E$ and hardness $H$ vs. VEC of B1-structured pseudobinary TMC/Ns (data taken from Ref.[4]). The scatter in $E$ and $H$ values of isoVEC alloys with VEC = 10.5 is highlighted using dashed orange ellipses.

Recent DFT calculations performed on a wider range of pseudobinary TMC/Ns (see Figure 1 data),[4, 19, 57] predicted that: B1-structure is stable when the VEC is below 10.6, *hardness decreases while ductility increases with increasing VEC*, and maximal toughness is expected for compounds with VECs between 9.5 and 10.5.

While VEC serves as a good initial indicator of mechanical properties, it alone is not sufficient to explain and predict the mechanical behavior of TMC/N compounds, where the primary bonds are non-metallic. Consider, for example, TMC/N alloys with the *same* VEC (hereto referred to as "isoVEC") of 10.5, which may be attained with group 5+6 nitrides [e.g., $V_{0.5}Cr_{0.5}N$, $Ta_{0.5}W_{0.5}N$, etc.] or group 6 carbonitrides [e.g., $CrC_{0.5}N_{0.5}$, $MoC_{0.5}N_{0.5}$, & $WC_{0.5}N_{0.5}$], see highlighted data in Figure 1. Worthy of note are the scattered values of the hardness $H$ and elastic moduli $E$ for the different isoVEC alloys. $H$ values, vary by over a factor of 3 from ~2 GPa for $Ta_{0.5}W_{0.5}N$ to ~7 GPa for $CrC_{0.5}N_{0.5}$ while the $E$ values vary two-fold from ~150 GPa to ~300 GPa for the same set of compounds. That is, mechanical properties of TMC/Ns with *the same VEC* vary with cation and anion composition and are different for unrelated TMN/C alloys that share the same VEC.

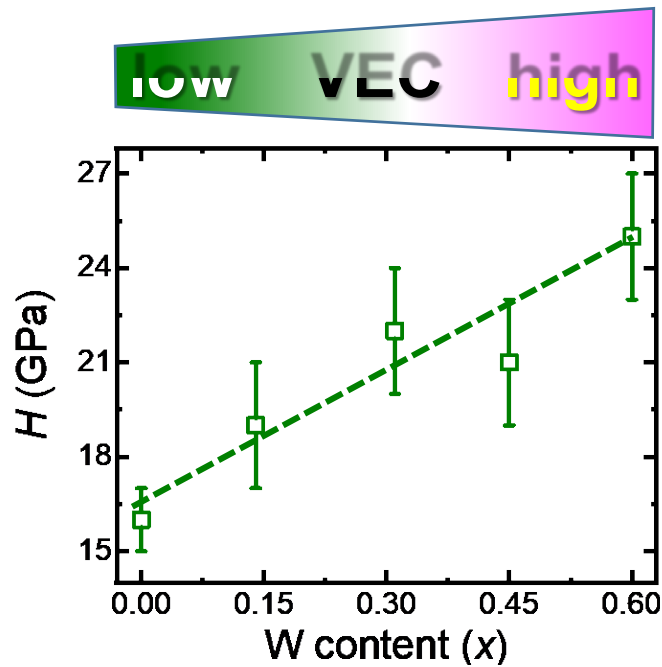

**Figure 2.** $V_{1-x}W_xN_y$(001) film hardness $H$ vs. $x$ [adapted from Ref. 54]. The increase in $H$ with increasing W content $x$, i.e. increasing VEC is in direct contrast with the DFT predicted trend of hardness vs VEC in Ref. 4.

Furthermore, the trends predicted by DFT do not always seem to agree with the experimentally observed behavior, see for example, data in Figure 2 from Refs. [13, 61] The observed increase in hardness of $V_{1-x}W_xN$(001) alloy films with increasing VEC (due to increasing W content) is not consistent with the DFT predicted trend. Similar inconsistencies between experimental results and predictions have also been observed in $V_{1-x}Mo_xN$(001) alloys[61] and group 4 carbonitrides.[62]

As justified below, research in the field in the near future will have to focus on the following question: given a set of transition-metal carbides and/or nitrides *with the same valence electron concentration* (isoVEC alloys or compounds), can we fundamentally understand their properties (e.g., strength, ductility) based on their cation and anion compositions?

An alternative approach to understanding mechanical behavior of a material is to look at its structural stability. In any structural alloys, including high-entropy alloys, the increase in ductility and accompanying decrease in strength with increasing VEC are attributed to change in the material's crystal structure from body centered cubic (*bcc*) to face-centered cubic (*fcc*).[63] In TMC/N, alloying can lead to the coexistence of several energetically equivalent phases, which can suppress dislocation motion and , thus, increase the material's strength.[26] For 4*d* TM nitrides, such multiphase structures are expected to form at VECs around 9.6.[27] *Clearly, it is necessary to go beyond simple descriptors such as VEC to understand the mechanical behavior in this class of materials with a mixture of ionic, covalent, and metallic bonding.* We suggest below several approaches aimed at the development of beyond-VEC models for understanding the mechanical properties of B1-phase TMC/Ns, specifically those with *isoVEC*.

As a step towards a fundamental understanding of what controls the mechanical properties of rocksalt TMC/Ns that have the same value of VEC, one can start investigating "easy" parameters of the anions and cations that make up the alloy's composition. These easy parameters (also called features in machine learning lingo) are usually computed from tabulated quantities specific to each anion or cation, and can be readily considered as potential descriptors of most properties of any complex alloys (not necessarily rocksalt-structured), ranging from the stability of single phases[64] to the catalytic activity.[65] From the ionic radii, atomic radii, electronegativities, Lewis acid strengths, lattice constants, cell volumes, etc. associated with each cation or with each parent structure (end member) that makes up the TMC/N alloy, we can derive various quantities associated with the alloy,[66] for example average and standard deviation for ionic radius of the cations; ionic radius for the anions; electronegativity of anions; electronegativity of cations; cell volume of end members, etc. We illustrate in Figure 3 some of these descriptors which have recently been tested for predicting the stability of single-phases of TMCs with multiple cations.[64]

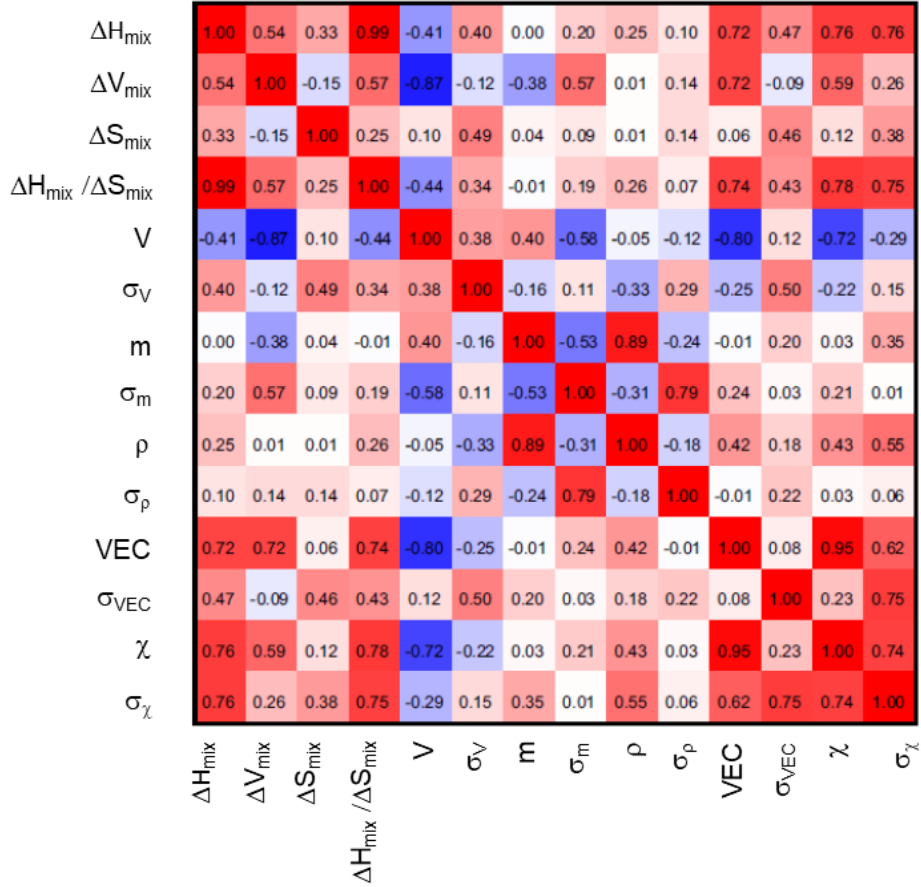

**Figure 3:** Feature selection for designing single-phase high-entropy carbides (From Ref. [64]). The values in the squares are the Pearson correlations, and they are listed so that highly-correlated features can be avoided when establishing machine learning models.

A closer look Figure 3 shows that VEC is strongly correlated with the average electronegativity $\bar{\chi}$ of the TMC constituents (end-members), and that the standard deviation of the VEC of constituents is also correlated reasonably well with the standard deviation of the electronegativities, $\sigma_\chi$. These observations suggest us that for isoVEC TMC/Ns, the discriminant factor that leads to the spread in mechanical properties (Figure 1) can very likely be related to the standard deviation of VEC, standard deviation of electronegativity, or some other measure of the variation of VEC or variation of electronegativity of the constituent TMC/Ns. While the purpose of this perspective article is merely to point out possible ways in which the use of VEC can be enhanced by looking at other descriptors of mechanical properties (as opposed to computing the performance of these descriptors), in Figure 4 we plot the hardness of TMC/Ns as computed by Balasubramanian et al.[4] versus $\Delta\bar{\chi}$, the difference in the average electronegativities of the cations and anions of the constituent TMC/Ns. As shown in Figure 4, a reasonable case can be made that for isoVEC TMC/N alloys, the hardness is controlled by $\Delta\bar{\chi}$.

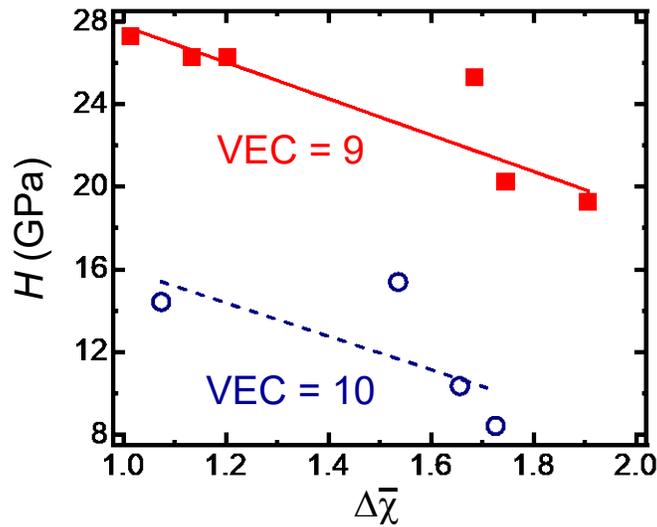

**Figure 4.** Correlation of isoVEC hardness $H$ with electronegativity difference, $\Delta\bar{\chi}$. The hardness data is from Ref. 4, while $\Delta\bar{\chi}$ is defined as the difference between the weighted average of the cation and anion electronegativities.

It is our hope that many other spread-type of descriptors (i.e., descriptors that highlight the differences between the constituent TMC/Ns, rather than average over them) would be tested in the near future. As shown in Figure 4, those are likely to account for the variation of properties for isoVEC TMC/Ns, with the caveat that perhaps different spread-type descriptors would perform differently for various mechanical properties. Before machine learning approaches, there were certain empirical formulations that showed significant value in designing hard materials. For example, Simunek[67] designed a semiempirical model based on coordination numbers, interatomic distances, valence electrons, volume containing valence electrons, etc. of each atom in the solid, and proposed a bond strength model and a hardness model based on these quantities. The approach[67] was not only simple and insightful, but also remarkably accurate, and was used by others to propose a range of super tough materials for further investigations at the level of DFT.[57] Therefore, in addition to the use of spread-type descriptors mentioned above, we encourage the exploration of semiempirical formulations for mechanical properties other than hardness, in particular for the Pugh's ratio, which can be used to classify the brittle versus ductile behavior of materials.[68]

Lastly, we posit as a central hypothesis that the mechanical behavior of transition-metal carbides and nitrides can be understood based on the *specific contributions of the VEC to different types of bonding*: ionic, covalent, and metallic as distinguishing factors between alloys with same VEC but different compositions and mechanical properties. Similar to the concept of the spread-type of descriptors discussed above, it is intuitively reasonable to expect that for isoVEC materials the partition of the valence electrons between ionic, covalent, and metallic bonding becomes important. However, this partition is not of the "easy" type described above, in the sense that it cannot be derived from tabulated properties of cations or anions. Rather, it has to be computed from DFT calculations, and we outline below a tentative procedure for such determination; improvements, validation, and testing would hopefully come from future work.

In refractory ceramics, the primary bonds are not metallic, as they occur between the TM and the C (or N) atom. In other words, it is not just about how many electrons are packed in a unit cell, but also how exactly they contribute to bonding and to mechanical properties. A TM atom is *not* surrounded by other 12 TM atoms (as it would in an fcc metal), but rather it is coordinated with 6 carbon or nitrogen atoms with which it exchanges or shares electrons. To illustrate the difference between fcc metals and B1 structured carbides/nitrides more concretely, we use the electron localization function (ELF),[69, 70] which represents the probability of finding a pair of electrons in the same region in space, and is related to the electron density and the curvature of the electron-electron correlation function.[71, 72]

While ELF is different than the electron density $n(\mathbf{r})$, it qualitatively exploits and amplifies spatial variations of $n$ to provide a clearer picture of the bonding. For example, in covalent materials such as silicon, ELF shows maxima between atoms, in metals it shows maxima in the interstitial sites, while in ionic compounds localization occurs mainly on the anion.

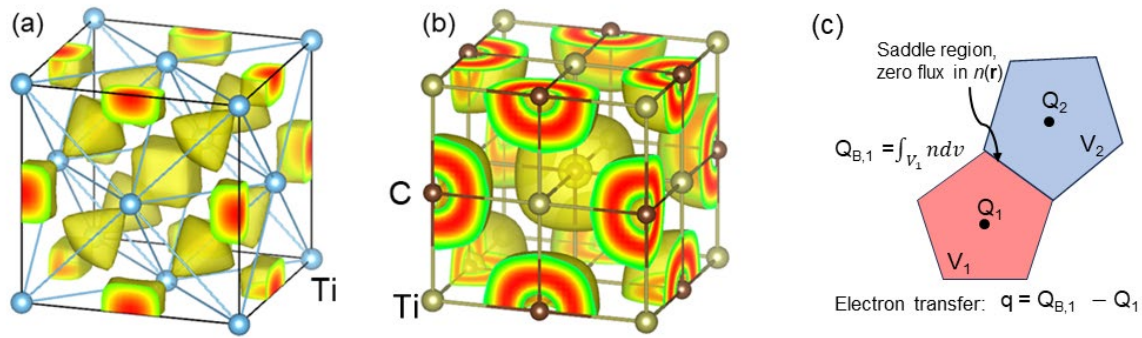

**Figure 5.** Electron localization function (ELF, isosurfaces) for (a) pure metal Ti in fcc structure and (b) for a transition-metal carbide, TaC in the B1 structure. For the elemental Ti, the bond is metallic, with no net charge on any atom and with electron localization primarily in the interstitial sites. The fcc Ti structure was chosen for ease of site comparison with TiC, but the hcp ground state of Ti also shows localization of electrons in the interstitial sites. For TiC in (b), the electron localization occurs around the C atom. (c) Schematic view of the Bader volumes around two atoms (black dots) in the lattice, with the number of electrons evaluated over these volumes.

Figure 5 shows the DFT-calculated ELF for a metal (Ti) and a transition-metal carbide (TMC), B1-structured TiC. For metallic Ti, each atom has 12 other Ti atoms surrounding it (Figure 5a), while for B1-structured TiC in Figure 5b — other rocksalt TMC/Ns are similar— each Ti atom has 6 C atoms as nearest neighbors (NN) and 12 Ti atoms as next-nearest neighbors (NNN). As a result, the ELFs for the two cases are vastly different. All atoms in the metal are neutral, while in the TMC, the C acquired electrons and has a net negative charge; necessarily, the TM atom in TMC will have a positive charge equal in absolute value to that on the C atom.

Figure 5(c) shows how electron charges can be "ascribed" to an atom $i$ using the Bader approach,[73] in which a volume $V_i$ is constructed for atom $i$ based on planar boundaries placed at "zero flux" regions, i.e. at the saddle points of the electron density $n(\mathbf{r})$. Such boundaries surround each ion with a so-called Bader volume, leading to a clear, unambiguous way to assign charges to the atoms/ions in a solid (Figure 5c). As stated, we aim to partition the total number of electrons in the cell, VEC, into electrons that contribute to the three different types of bonding

$$\text{VEC} = N_{ionic} + N_{covalent} + N_{metallic},$$

where $N_{ionic}$ (or covalent, or metallic) is the number of electrons that are committed to ionic (or covalent, or metallic) bonds. From the Bader procedure,[73] we readily obtain $N_{ionic} = q$, the electron transfer from TM to C (or N). $N_{covalent}$ and $N_{metallic}$ have distinctive signatures in ELF plots (Figure 5a,b). ELF plots of electrons in covalent bonds show maxima exactly between two covalently bonded atoms, and the electrons contributing to metallic bonding show ELF maxima in the interstitial sites formed between 4 atoms (Figure 5a). In order to defined $N_{metallic}$ in TMC/Ns, we would have to compute the electronic charge located inside the smallest polyhedra with the TM at corners, or more specifically, inside the (maximal) spheres inscribed in the tetrahedra created by TM atoms in the B1 lattice (these are NNN). Such choice leaves the edges of the these tetrahedra to be surrounded by covalent electron distributions associated with the TM-C or TM-N bonds. With this definition of $N_{metallic}$, the remaining balance of VEC will be the number of electrons committed to the covalent bond, $N_{covalent} = VEC - q - N_{metallic}$.

This partition procedure naturally offers now several descriptors beyond VEC: one can use $(q, N_{covalent}, N_{metallic})$ as independent descriptors, or use $(VEC, q, N_{metallic})$. In other words, at the same VEC, the parameters that lead to different mechanical and structural properties would be $q$ and $N_{metallic}$, both of which computable from the Bader charge analysis. This procedure differs from past attempts[74] to separate the electronic density into bonds (bond partition charge) in that it does not require empirical parameterizations of the overlap regions between the atoms. Any shortcoming of the bond-specific electron partitioning described above would also be a shortcoming of the Bader method itself. It is our hope that this procedure to discriminate between the electrons associated to different types of bonding will enable the use of (VEC, $q$, $N_{metallic}$) as independent variables with respect to which to analyze most mechanical properties, e.g., Young's modulus, bulk modulus, shear modulus, hardness, Pugh's ratio, Poisson ratio, etc. In particular, the (VEC, $q$, $N_{metallic}$) analysis of the Pugh's ratio and Poisson ratio could emerge as a promising avenue to discover hard and ductile TMC/Ns alloys.

To summarize, the quest for appropriate descriptors of mechanical properties in B1-structured refractory carbides and their alloys is long standing problem, both partially alleviated and confounded by the use of VEC as a predictor of mechanical behavior. While providing a (necessarily incomplete) review of the approaches to understand and design hard and tough materials, we proposed that spread-type of descriptors can be used to understand the different mechanical properties of isoVEC TMC/Ns and illustrated the correlation between hardness and electronegativity difference between cations and anions. Furthermore, we proposed a way to ascertain the electron "content" responsible for the ionic, metallic, and covalent bonding, based on available tools in analysis of charge distribution. This partitioning between types of bonding is unambiguous, parameter-free, and as such will be usable to bonding in other crystalline structures that are mechanically stable (not necessarily ground states), in particular carbides, nitrides, oxides, oxycarbides in which bonding is not solely ionic. The partitioning of VEC between the different types of bonds is particularly useful for simple structures such as rocksalt, as there only a few types of cation-anion and (possibly) cation-cation second neighbor bonds.

The development and validation of relationships between simple bonding descriptors beyond VEC and mechanical behavior will shed light on the physical origins of alloying for creating multiple and potentially opposing functionalities (e.g., strength and ductility) in refractory carbide or nitride alloys. Given that this class of materials, cubic transition-metal compounds, are of the form MX, where X = C, N, O, or other elements with equi-atomic cations and anions, we expect that the set of descriptors outlined here should be applicable to assess the stability and mechanical behavior of multi-cation as well as multi-anion alloys, including high-entropy alloys. The ability to *predict* and hence design materials with properties of interest for certain applications, and the evaluation of the fundamental limits or ranges of mechanical properties for use in various applications has been a major quest for materials scientists. The approaches presented in this perspective can extend our current knowledge and provide new insights into the role of alloying on the mechanical behavior of TMC/Ns with iso-valent cation composition.

**Acknowledgments**

We acknowledge the support from the Air Force Office of Scientific Research (AFOSR, Dr. Ali Sayir) Grant Nos. FA9550-18-1-0050 and FA9550-20-1-0184.